\documentclass[conference]{IEEEtran}
\IEEEoverridecommandlockouts
\usepackage{cite}
\usepackage{amsmath,amssymb,amsfonts,upgreek}
\usepackage{algorithmic}
\usepackage{algorithm}
\usepackage{graphicx}
\usepackage{textcomp}
\usepackage{xcolor}
\usepackage{stfloats}

\usepackage{hhline}
\usepackage{makecell}
\usepackage{booktabs}
\usepackage[flushleft]{threeparttable}
\usepackage{balance}

\DeclareMathOperator{\diag}{diag}

\def\BibTeX{{\rm B\kern-.05em{\sc i\kern-.025em b}\kern-.08em
    T\kern-.1667em\lower.7ex\hbox{E}\kern-.125emX}}
\begin{document}

\title{Deep Complex-valued Radial Basis Function \\ Neural Networks and Parameter Selection}


\author{\IEEEauthorblockN{Jonathan A. Soares, Vinícius H. Luiz, Dalton S. Arantes, and Kayol S. Mayer}
\IEEEauthorblockA{\textit{Department of Communications, School of Electrical and Computer Engineering} \\
\textit{Universidade Estadual de Campinas -- UNICAMP}, Campinas, SP, Brazil\\
j229966@dac.unicamp.br, v245342@dac.unicamp.br, dalton@unicamp.br, kayol@unicamp.br}
}

\maketitle

\begin{abstract}
In the ever-evolving field of artificial neural networks and learning systems, complex-valued neural networks (CVNNs) have become a cornerstone, achieving exceptional performance in image processing and telecommunications. 
More precisely, in digital communication systems, CVNNs have been delivering significant results in tasks like equalization, channel estimation, beamforming, and decoding. Among the CVNN architectures, the complex-valued radial basis function neural network (C-RBF) stands out, especially when operating in noisy environments such as 5G multiple-input multiple-output~(MIMO) systems. In such a context, this paper extends the classical shallow C-RBF to deep architectures, increasing its flexibility for a wider range of applications. Also, based on the parameter selection of the phase transmittance radial basis function (PT-RBF) neural network, we propose an initialization scheme for the deep C-RBF. Via rigorous simulations conforming to 3GPP TS 38 standards for digital communications, our method not only outperforms conventional initialization strategies like random, $K$-means, and constellation-based methods but it also seems to be the only approach to achieve successful convergence for deep C-RBF architectures. These findings pave the way to more robust and efficient neural network deployments in complex-valued digital communication systems.

\end{abstract}

\section{Introduction}

Recently, in communication systems, complex-valued neural networks~(CVNNs) have been studied in several applications, such as equalization, channel estimation, beamforming, and decoding~\cite{Mayer2019a,Ding2020,Zhang2021b,Li2022,Mayer2020a,Kamiyama2021,Freire2020,Soares2021,Xu2022, Chu2022,Mayer2022,Yang2022b,Xiao2023}. This growing interest is related to enhanced functionality, improved performance, and reduced training time when compared with real-valued neural networks~(RVNNs)~\cite{Hirose2012b, Barrachina2021, Cruz2022, Zhang2022}.

The effectiveness of neural networks is critically dependent on several factors, such as initialization, regularization, and optimization~\cite{Humbird2019}. Although regularization and optimization techniques are vital to speed up the training process and reduce steady-state error~\cite{Hu2021}, depending on the initial parameter selection the neural network can get stuck at local minima, achieving suboptimal solutions~\cite{Narkhede2022}. For radial basis function~(RBF)-based neural networks, this problem is even worse since, for each layer, there are four parameters~(synaptic weight, bias, center vectors, and center variances) in contrast to two parameters~(synaptic weights and bias) of usual multilayer perceptron neural networks.

In this context, with a focus on the complex-valued radial basis function~(C-RBF) neural network~\cite{Mayer2022}, we propose an extension for deep learning and a novel parameter selection scheme. This scheme aims to initialize synaptic weights, biases, center vectors, and center variances in the complex domain. Notably, existing literature offers limited guidance on initialization techniques for multilayer RBF-based CVNNs. Despite this gap, our study compares the proposed approach against well-known methods such as random initialization~\cite{Wallace2005}, $K$-means clustering~\cite{Turnbull2005}, and constellation-based initialization~\cite{Loss2007}. To the best of our knowledge, this is the first work proposing the architecture, training algorithm, and parameter selection for a multi-layered C-RBF.


\section{C-RBF Neural Networks}
\label{sec:cvnns}

The complex-valued Gaussian neuron is a natural extension of the well-known Gaussian neuron for the complex domain \cite{Chen1994}. Similarly to its real-valued version, the output of the C-RBF neuron is described as
\begin{equation}
\label{eq:gaussianNeuronCRBF}
    y[n]=w[n]\phi[n]+b[n],
\end{equation}
in which $\phi[n]\in \mathbb{R}$ is the Gaussian kernel output
\begin{equation}
\label{eq:gaussianNeuronCRBF_phi}
   \phi[n]=\exp\left(-\frac{\left \lVert\mathbf{x}[n]-\boldsymbol{\upgamma}[n]\right \rVert_2^2}{\sigma[n]} \right ),
\end{equation}
and $\boldsymbol{\upgamma}[n]\in\mathbb{C}^P$ is the Gaussian center, $\sigma[n]\in \mathbb{R}$ is the variance. Note that the bias $b[n]\in \mathbb{C}$ is a linear complex-valued synaptic weight like $w[n]\in \mathbb{C}$, but considering the Gaussian output equals one. Unlike the RBF neuron, the C-RBF neuron Gaussian center, synaptic weight, and bias are complex-valued free parameters, which are essential to map a complex-valued input $\mathbf{x}[n]\in\mathbb{C}^P$ into a complex-valued output $y[n]\in \mathbb{C}$. By \eqref{eq:gaussianNeuronCRBF_phi}, the complex-valued input is firstly mapped into a real-valued scalar via the Euclidean norm of the Gaussian kernel. As the variance is also a real-valued parameter, the Gaussian kernel output is consequently a real-valued scalar. Thus, the complex mapping to the output is only possible because of the synaptic weights and bias. 

\subsection{Shallow C-RBF}

The main differences between the C-RBF and RBF regard the free parameters domains and the backpropagation training. In a C-RBF neural network with $P$ inputs, $R$ outputs, and $M$ Gaussian neurons, the vector of outputs $\mathbf{y}[n] \in \mathbb{C}^R$ is given by
\begin{equation}
\label{eq:CRBF_output}
    \mathbf{y}[n]=\mathbf{W}[n]\boldsymbol{\upphi}[n]+\mathbf{b}[n], 
\end{equation}
where $\mathbf{W}[n]\in\mathbb{C}^{R \times M}$ is the matrix of synaptic weights, $\boldsymbol{\upphi}[n]\in \mathbb{R}^M$ is the vector of Gaussian kernels, and $\mathbf{b}[n]\in \mathbb{C}^R$ is the vector of bias.

The $m$-th Gaussian kernel of $\boldsymbol{\upphi}[n]$ is formulated as
\begin{equation}
\label{eq:CRBF_GaussianKernel}
\phi_m[n]=\exp\left(-\frac{\left \lVert\mathbf{x}[n]-\boldsymbol{\upgamma}_m[n]\right \rVert_2^2}{\sigma_m[n]} \right ),
\end{equation}
in which $\boldsymbol{\upgamma}_m[n]\in \mathbb{C}^P$ and $\sigma_m[n]\in\mathbb{R}$ are the $m$-th vectors of Gaussian Centers and variances, respectively.

Albeit $\mathbf{b}[n]$ and $\mathbf{W}[n]$ can be considered as only one free parameter, for the sake of simplicity we assume it as separate free parameters. Therefore, the C-RBF is a shallow ANN with four free parameters (i.e., matrix of synaptic weights $\mathbf{W}[n]$; vector of bias $\mathbf{b}[n]$; matrix of Gaussian centers $\boldsymbol{\Gamma}[n]\in \mathbb{C}^{M \times P}$; and vector of variances $\boldsymbol{\upsigma}[n]\in \mathbb{R}^M$), whose updates are performed via the steepest descent algorithm as
\begin{equation} \label{eq:CRBF_steepest_descent}
\begin{gathered}
w_{r,m}[n+1] = w_{r,m}[n]-\eta_w \nabla_w J[n],\\
 b_{r}[n+1] = b_{r}[n]-\eta_b \nabla_b J[n],\\
 \boldsymbol{\upgamma}_{m}[n+1] = \boldsymbol{\upgamma}_{m}[n]-\eta_\gamma \nabla_\gamma J[n],\\
  \sigma_{m}[n+1]  = \sigma_{m}[n]-\eta_\sigma \nabla_\sigma J[n],
\end{gathered}
\end{equation}
where $\eta_w$, $\eta_b$, $\eta_\gamma$, and $\eta_\sigma$ are the respective adaptive steps of $w_{r,m}$, $b_r$, $\boldsymbol{\upgamma}_{m}$, and $\sigma_{m}$. Also, $\nabla_w$, $\nabla_b$, $\nabla_\gamma$, and $\nabla_\sigma$ are the complex gradient operators of $w_{r,m}$, $b_{r}$, $\boldsymbol{\upgamma}_{m}$, and $\sigma_{m}$, respectively. Furthermore, $J[n]$ is the quadratic cost function
\begin{equation} 
\label{eq:GeneralizedDeltaRuleCostFunctionCRBF}
    J[n] = \frac{1}{2}\left \lVert\mathbf{e}[n]\right \rVert_2^2=\frac{1}{2}\left \lVert \mathbf{d}[n]-\mathbf{y}[n] \right \rVert_2^2,
\end{equation}
in which $\mathbf{e}[n]=\mathbf{d}[n]-\mathbf{y}[n]\in\mathbb{C}^R$ is the error vector and $\mathbf{d}[n]\in\mathbb{
C}^R$ is the vector of desired outputs.

Solving the gradients in \eqref{eq:CRBF_steepest_descent}, and organizing the equations in matrix structures, we obtain
\begin{equation} 
\label{eq:CRBFupdateEquationsMatrix}
\begin{gathered}
    \mathbf{W}[n+1] = \mathbf{W}[n] + \eta_w \mathbf{e}[n]\boldsymbol{\upphi}^T[n],\\
    \mathbf{b}[n+1] = \mathbf{b}[n] + \eta_b \mathbf{e}[n],\\
    \boldsymbol{\Gamma}[n+1] = \boldsymbol{\Gamma}[n] + \eta_{\boldsymbol{\upgamma}} \mathrm{diag}\left(\boldsymbol{\xi}[n] \odot\boldsymbol{\upbeta}[n]\right )\left(\mathbf{X}[n]-\boldsymbol{\Gamma}[n]\right ),\\
    \boldsymbol{\upsigma}[n+1] = \boldsymbol{\upsigma}[n] + \eta_{\sigma}  \boldsymbol{\xi}[n]\odot\boldsymbol{\upbeta}[n]\odot\mathbf{v}[n],
\end{gathered}
\end{equation}
where $\boldsymbol{\xi}[n]=\Re\left(\mathbf{W}[n]\right)^T\Re\left(\mathbf{e}[n]\right)+\Im\left(\mathbf{W}[n]\right)^T\Im\left(\mathbf{e}[n]\right)\in\mathbb{R}^M$ is the vector of synaptic transmittance, and $\boldsymbol{\upbeta}[n]\in\mathbb{R}^{M}$ is the vector of Gaussian weighted kernel, with the $m$-th component $\beta_m[n]=\phi_m[n]/\sigma_m[n]$. The expanded input matrix $\mathbf{X}[n]\in \mathbb{C}^{M\times P}$ is
\begin{equation} 
\label{eq:CRBFinputMatrix}
    \mathbf{X}[n]=\begin{bmatrix}
                    \textbf{\text{---}} & \mathbf{x}^T[n] & \textbf{\text{---}}\\
                    \textbf{\text{---}} & \mathbf{x}^T[n] & \textbf{\text{---}}\\
                     & \vdots & \\
                     \textbf{\text{---}} & \mathbf{x}^T[n] & \textbf{\text{---}}
    \end{bmatrix}.
\end{equation} 

\begin{figure*}[!b]
\hrulefill

\normalsize
\newcounter{mytempeqncnt}
\setcounter{mytempeqncnt}{\value{equation}}
\setcounter{equation}{14}
\begin{equation} 
\label{eq:deepPTRBFAux1}
\boldsymbol{\uppsi}^{\{l\}}[n]=\begin{cases}\left[\mathbf{Y}^{\{l\}}[n]-\boldsymbol{\Upgamma}^{\{l+1\}}[n]\right]^T\boldsymbol{\updelta}^{\{l+1\}}[n]\mathbf{1_{I^{\{l+1\}}}},& \mathrm{for} \ 0<l<L,\\
\mathbf{d}[n]-\mathbf{y}[n]=\mathbf{e}[n],& \mathrm{for} \ l=L,
\end{cases}
\end{equation}
\begin{equation} 
\label{eq:deepPTRBFAux2}
\boldsymbol{\updelta}^{\{l\}}[n]=-\diag\left\{\left[\Re\left(\mathbf{W}^{\{l\}}[n]\right)^T\Re\left(\boldsymbol{\uppsi}^{\{l\}}[n]\right)+\Im\left(\mathbf{W}^{\{l\}}[n]\right)^T\Im\left(\boldsymbol{\uppsi}^{\{l\}}[n]\right)\right]\odot\boldsymbol{\upbeta}^{\{l\}}[n]\right\}.
\end{equation}
\setcounter{equation}{\value{mytempeqncnt}}
\end{figure*}

\subsection{Proposed Deep C-RBF}

The deep C-RBF is defined with $L$ hidden layers~(excluding the input layer), where the superscript $l \in [0,\,1,\,\cdots,\,L]$ denotes the layer index and $l=0$ is the input layer. The $l$-th layer~(excluding the input layer $l=0$) is composed by $I^{\{l\}}$ neurons, $O^{\{l\}}$ outputs, and has a matrix of synaptic weights $\mathbf{W}^{\{l\}}\in\mathbb{C}^{O^{\{l\}}\times I^{\{l\}}}$, a bias vector $\mathbf{b}^{\{l\}}\in\mathbb{C}^{O^{\{l\}}}$, a matrix of center vectors $\boldsymbol{\Gamma}^{\{l\}}\in\mathbb{C}^{I^{\{l\}}\times O^{\{l-1\}}}$, and a variance vector $\boldsymbol{\upsigma}^{\{l\}}\in\mathbb{R}^{I^{\{l\}}}$. Notice that $\mathbf{\bar{x}}\in \mathbb{C}^P$ is the deep C-RBF normalized input vector~($P$ inputs) and $\mathbf{y}^{\{L\}}\in \mathbb{C}^R$ is the deep C-RBF output vector~($R$ outputs). The $l$-th hidden layer output vector $\mathbf{y}^{\{l\}} \in \mathbb{C}^{O^{\{l\}}}$ is given by
\begin{equation}
\label{eq:deepPTRBFNN_output}
    \mathbf{y}^{\{l\}}=\mathbf{W}^{\{l\}}\boldsymbol{\upphi}^{\{l\}}+\mathbf{b}^{\{l\}}, 
\end{equation}
where $\boldsymbol{\upphi}^{\{l\}}\in \mathbb{R}^{I^{\{l\}}}$ is the vector of Gaussian kernels.

The $m$-th Gaussian kernel of the $l$-th hidden layer is formulated as
\begin{equation}
\label{eq:kernel}
\phi_m^{\{l\}}=\exp\left[-v_m^{\{l\}}\right],      
\end{equation}
in which $v_m^{\{l\}}$ is the $m$-th Gaussian kernel input of the $l$-th hidden layer, described as
\begin{equation}
\label{eq:kernel_argument}
v_m^{\{l\}}=\frac{\left \Vert\mathbf{y}^{\{l-1\}}-\boldsymbol{\upgamma}_m^{\{l\}}\right \Vert_2^2}{\sigma_m^{\{l\}}},
\end{equation}
where $\mathbf{y}^{\{l-1\}}\in \mathbb{C}^{O^{\{l-1\}}}$ is the output vector of the $(l-1)$-th hidden layer~(except for the first hidden layer that $\mathbf{y}^{\{0\}}=\mathbf{\bar{x}}$), $\boldsymbol{\upgamma}_m^{\{l\}}\in \mathbb{C}^{O^{\{l-1\}}}$ is the $m$-th vector of Gaussian centers of the $l$-th hidden layer, $\sigma_m^{\{l\}}\in\mathbb{R}$ is the respective $m$-th variance.

Similarly to \eqref{eq:CRBF_steepest_descent}, we can define a generalized steepest descent algorithm to the $l$-th layer, as
\begin{equation}
\label{eq:PTRBFDNN_steepest_descent_decoder}
\begin{gathered}
w^{\{l\}}_{r,m}[n+1] = w^{\{l\}}_{r,m}[n]-\eta_w \nabla^{\{l\}}_w J[n],\\
 b^{\{l\}}_{r}[n+1] = b^{\{l\}}_{r}[n]-\eta_b \nabla^{\{l\}}_b J[n],\\
 \boldsymbol{\upgamma}^{\{l\}}_{m}[n+1] = \boldsymbol{\upgamma}^{\{l\}}_{m}[n]-\eta_\gamma \nabla^{\{l\}}_\gamma J[n],\\
  \sigma^{\{l\}}_{m}[n+1]  = \sigma^{\{l\}}_{m}[n]-\eta_\sigma \nabla^{\{l\}}_\sigma J[n].
\end{gathered}
\end{equation}

Solving the gradients in \eqref{eq:PTRBFDNN_steepest_descent_decoder}, and organizing the resulting equations in matrix structures, we obtain
\begin{equation} 
\label{eq:deepPTRBFupdateMatrices}
\begin{gathered}
    \mathbf{W}^{\{l\}}[n+1] = \mathbf{W}^{\{l\}}[n] + \eta_w^{\{l\}} \boldsymbol{\uppsi}^{\{l\}}[n]\left(\boldsymbol{\upphi}^{\{l\}}[n]\right)^T,\\
    \mathbf{b}^{\{l\}}[n+1] = \mathbf{b}^{\{l\}}[n] + \eta_b^{\{l\}} \boldsymbol{\uppsi}^{\{l\}}[n],\\
    \boldsymbol{\Gamma}^{\{l\}}[n+1] = \boldsymbol{\Gamma}^{\{l\}}[n] - \eta_{\boldsymbol{\upgamma}}^{\{l\}}  \boldsymbol{\updelta}^{\{l\}}[n]\left(\mathbf{Y}^{\{l-1\}}[n]-\boldsymbol{\Upgamma}^{\{l\}}[n]\right),\\
    \boldsymbol{\upsigma}^{\{l\}}[n+1] = \boldsymbol{\upsigma}^{\{l\}}[n] -\eta_{\sigma}^{\{l\}}\boldsymbol{\updelta}^{\{l\}}[n]\mathbf{v}^{\{l\}}[n],
\end{gathered}
\end{equation}
where $\boldsymbol{\uppsi}^{\{l\}}$ and $\boldsymbol{\updelta}^{\{l\}}[n]$ are presented at the bottom of the next page, and $\boldsymbol{\upbeta}^{\{l\}}[n]\in\mathbb{R}^{I^{\{l\}}}$ is the vector of Gaussian weighted kernel of the $l$-th hidden layer, with the $m$-th component $\beta_m^{\{l\}}[n]=\phi_m^{\{l\}}[n]/\sigma_m^{\{l\}}[n]$. The vector $\mathbf{1_{I^{\{l+1\}}}}$ is composed of $I^{\{l+1\}}$ ones, and $\mathbf{Y}^{\{l\}}[n]\in \mathbb{C}^{I^{\{l+1\}}\times O^{\{l\}}}$ is the expanded matrix of layer's outputs, in which each row is given by
\begin{equation} 
\label{eq:deepPTRBFbottleneckMatrix}
    \mathbf{Y}^{\{l\}}[n]=\begin{bmatrix}
                    \textbf{\text{---}} & \left(\mathbf{y}^{\{l\}}[n]\right)^T & \textbf{\text{---}}\\
                    \textbf{\text{---}} & \left(\mathbf{y}^{\{l\}}[n]\right)^T & \textbf{\text{---}}\\
                     & \vdots & \\
                     \textbf{\text{---}} & \left(\mathbf{y}^{\{l\}}[n]\right)^T & \textbf{\text{---}}
    \end{bmatrix}.
\end{equation} 
 
\section{Proposed deep C-RBF parameter initialization}
\label{sec:model}
\setcounter{equation}{16}

Based on \cite{Soares2024}, to properly initialize the deep C-RBF parameters, we first need to understand the relationship between the input vector $\mathbf{x}$ and the Gaussian center vectors $\boldsymbol{\upgamma}_m^{\{1\}}$. In \eqref{eq:kernel}, regarding \eqref{eq:kernel_argument}, and keeping $\sigma_m^{\{1\}}$ constant, the closer $\boldsymbol{\upgamma}_m^{\{1\}}$ is to $\mathbf{x}$, higher is the value of $\phi_m^{\{1\}}$. For example, if $\boldsymbol{\upgamma}_m^{\{1\}}=\mathbf{x}$ then $\phi_m^{\{1\}} = 1$. On the other hand, if $\boldsymbol{\upgamma}_m^{\{1\}}$ is set far from $\mathbf{x}$, then $\phi_m^{\{1\}}\to 0$. In this context, to not saturate or vanish $\phi_m^{\{1\}}$, we assume $\mu_\mathbf{\bar{x}}=\mu_{\boldsymbol{\upgamma}^{\{1\}}} = 0$ and $\sigma^2_\mathbf{\bar{x}}=\sigma^2_{\boldsymbol{\upgamma}^{\{1\}}}$, where $\mathbf{\bar{x}}$ is the normalized input dataset. Furthermore, we expect that depending on the dataset inputs, $\phi_m^{\{1\}}$ varies reasonably. For example, considering $v_m^{\{1\}}=5$ and $v_m^{\{1\}}=10$, the variation in $\phi_m^{\{1\}}$ is only $4.54\times10^{-5}$. In contrast, considering $v_m^{\{1\}}=0$ and $v_m^{\{1\}}=3$, the variation in $\phi_m^{\{1\}}$ is $0.95$. Then, it is desirable that $\mu_{v^{\{1\}}}$ is not too large. Based on Appendix A of \cite{Soares2024}, the expected value of $v^{\{1\}}$ is
\begin{equation}
\label{eq:mean_v_1}
   \mu_{\mathbf{v}^{\{1\}}} = \frac{P}{c_\sigma}\left[\sigma^2_{\Re\left(\mathbf{\bar{x}}\right)}+\sigma^2_{\Im\left(\mathbf{\bar{x}}\right)}
   +\sigma^2_{\Re\left(\boldsymbol{\upgamma}_m^{\{1\}}\right)}+\sigma^2_{\Im\left(\boldsymbol{\upgamma}_m^{\{1\}}\right)}\right],
\end{equation}
in which $\sigma^2_{\mathbf{\bar{x}}}=2\sigma^2_{\Re\left(\mathbf{\bar{x}}\right)}=2\sigma^2_{\Im\left(\mathbf{\bar{x}}\right)}$ is the variance of $\mathbf{\bar{x}}$, $\sigma^2_{\boldsymbol{\upgamma}_m^{\{1\}}}=2\sigma^2_{\Re\left(\boldsymbol{\upgamma}_m^{\{1\}}\right)}=2\sigma^2_{\Im\left(\boldsymbol{\upgamma}_m^{\{1\}}\right)}$ is the variance of $\boldsymbol{\upgamma}_m^{\{1\}}$, $c_\sigma=\sigma_m^{\{1\}}\,\,\forall \,\,m$, and $c_\sigma$ is a positive and nonzero constant.

As $\sigma^2_\mathbf{\bar{x}}=\sigma^2_{\boldsymbol{\upgamma}^{\{1\}}}$, from \eqref{eq:mean_v_1}, we have
\begin{equation}
\label{eq:var_x}
    \sigma^2_\mathbf{\bar{x}} = \frac{c_\sigma\mu_{\mathbf{v}^{\{1\}}}}{2P}.
\end{equation}

Based on \eqref{eq:var_x}, the normalized input is given as
\begin{equation}
\label{eq:norm_x}
    \mathbf{\bar{x}} = \frac{(\mathbf{x}-\mu_{\mathbf{x}})}{\sqrt{\sigma^2_{\mathbf{x}}}} \sqrt{\frac{c_\sigma \mu_{\mathbf{v}^{\{1\}}}}{2P}},
\end{equation}
where $\mu_{\mathbf{x}}$ and $\sigma^2_{\mathbf{x}}$ are applied to adjust the mean and variance of $\mathbf{\bar{x}}$ before the normalization by \eqref{eq:var_x}.

Similarly, in the first hidden layer, the normalized matrix of center vectors is
\begin{equation}
\label{eq:gamma_init}
    \boldsymbol{\Gamma}^{\{1\}} \sim \mathcal{C}\mathcal{G}\left(0,\frac{c_\sigma \mu_{\mathbf{v}^{\{1\}}}}{2P} \right).
\end{equation}

In order to normalize the output dataset $\mathbf{d}$, we need to compute the variance of the output vector $\mathbf{y}^{\{L\}}$, by
\begin{equation}
\label{eq:variance_y_output}
    \sigma^2_{\mathbf{y}^{\{L\}}} = \mathrm{Var}\left[\mathbf{W}^{\{L\}}\boldsymbol{\upphi}^{\{L\}}+\mathbf{b}^{\{L\}}\right].
\end{equation}

We assume that $\mathbf{b}^{\{l\}}$ is initialized with zeros, for all layers. Thus, based on Appendix~C of \cite{Soares2024}, \eqref{eq:variance_y_output} results in
\begin{equation}
        \sigma _{\mathbf{y}^{\{l\}}}^{2} = \frac{12}{5}\left( \frac{\sigma _{\upgamma _{m}^{\{l\}}}^{2}}{c_{\sigma}\exp\left({\mu_{\mathbf{v}^{\{l\}}}}\right)}\right)^{2} I^{\{l\}}O^{\{l-1\}} \sigma _{\mathbf{W}^{\{l\}}}^{2},
\end{equation}
where $\sigma^2_{\mathbf{W^{\{L\}}}}$ is the variance of $\mathbf{W}^{\{L\}}$, and $\mu_{\mathbf{v}^{\{L\}}}$ is the expected value of $\mathbf{v}^{\{L\}}$. Choosing $\sigma^2_{\mathbf{y}^{\{L\}}}=\sigma^2_\mathbf{\bar{d}}$, i.e., the variance of the C-RBF output equal to the variance of the normalized output dataset, yields the initialization of $\mathbf{W}^{\{L\}}$ as
\begin{equation}
\label{eq:w_init}
    \mathbf{W}^{\{L\}} \sim \mathcal{C}\mathcal{G}\left(0,\frac{\sigma^2_{\mathbf{\bar{d}}}}{\dfrac{12}{5}\left( \dfrac{\sigma _{\upgamma _{m}^{\{l\}}}^{2}}{c_{\sigma}\exp\left({\mu_{\mathbf{v}^{\{l\}}}}\right)}\right)^{2} \hspace{-2mm}I^{\{l\}} O^{\{l-1\}}} \right),
\end{equation}
in which the output dataset can be normalized by
\begin{equation}
\label{eq:norm_d}
    \mathbf{\bar{d}} = \frac{(\mathbf{d}-\mu_{\mathbf{d}})}{\sqrt{\sigma^2_{\mathbf{d}}}} \sqrt{\sigma^2_{\mathbf{y}^{\{L\}}}}=\frac{(\mathbf{d}-\mu_{\mathbf{d}})}{\sqrt{\sigma^2_{\mathbf{d}}}} \sqrt{\frac{c_\sigma\mu_{\mathbf{v}^{\{L\}}}}{2R}}.
\end{equation}

Relying on \eqref{eq:gamma_init}, we can generalize the initialization of $\boldsymbol{\Gamma}^{\{l\}}$, as
\begin{equation}
\label{eq:gamma_init_layers}
    \boldsymbol{\Gamma}^{\{l\}} \sim \mathcal{C}\mathcal{G}\left(0,\frac{c_\sigma \mu_{\mathbf{v}^{\{l\}}}}{2O^{\{l-1\}}} \right).
\end{equation}

From \eqref{eq:var_x}, the variance of the output hidden layers can be considered as
\begin{equation}
\label{eq:var_y_hidden}
    \sigma^2_{\mathbf{y}^{\{l\}}} = \frac{c_\sigma\mu_{\mathbf{v}^{\{l+1\}}}}{2O^{\{l\}}},
\end{equation}
where, replacing $\sigma^2_{\mathbf{\bar{d}}}$ by \eqref{eq:var_y_hidden} into \eqref{eq:w_init}, yields

\begin{equation}
\label{eq:w_init_layers}
    \mathbf{W}^{\{L\}} \sim \mathcal{CG}\left( 0,\frac{5c_{\sigma } O^{\{l-1\}}}{6I^{\{l\}} O^{\{l\}} \mu_{\mathbf{v}^{\{l+1\}}}\exp (-2\mu _{\mathbf{v}^{\{l\}}} )}\right).
\end{equation}

It is important to note that, \eqref{eq:norm_x} and \eqref{eq:norm_d} only hold for $\sigma^2_{\mathbf{x}}=2\sigma^2_{\Re\left(\mathbf{x}\right)}=2\sigma^2_{\Im\left(\mathbf{x}\right)}$ and $\sigma^2_{\mathbf{d}}=2\sigma^2_{\Re\left(\mathbf{d}\right)}=2\sigma^2_{\Im\left(\mathbf{d}\right)}$, respectively. For the particular case of $\sigma^2_{\Re\left(\mathbf{x}\right)}\neq\sigma^2_{\Im\left(\mathbf{x}\right)}$ and $\sigma^2_{\Re\left(\mathbf{d}\right)}\neq\sigma^2_{\Im\left(\mathbf{d}\right)}$, then \eqref{eq:norm_x} and \eqref{eq:norm_d} become
\begin{equation}
    \mathbf{\bar{x}} \hspace{-1mm}= \hspace{-1.2mm}\left[\frac{(\Re\left(\mathbf{x}\right)-\mu_{\Re\left(\mathbf{x}\right)})}{\sqrt{2\sigma^2_{\Re\left(\mathbf{x}\right)}}}+\jmath\frac{(\Im\left(\mathbf{x}\right)-\mu_{\Im\left(\mathbf{x}\right)})}{\sqrt{2\sigma^2_{\Im\left(\mathbf{x}\right)}}}\right]\hspace{-1.4mm}\sqrt{\frac{c_\sigma \mu_{\mathbf{v}^{\{1\}}}}{2P}},
\end{equation}
\begin{equation}
    \hspace{-1mm}\mathbf{\bar{d}} \hspace{-1mm}= \hspace{-1.2mm}\left[\frac{(\Re\left(\mathbf{d}\right)-\mu_{\Re\left(\mathbf{d}\right)})}{\sqrt{2\sigma^2_{\Re\left(\mathbf{d}\right)}}}+\jmath\frac{(\Im\left(\mathbf{d}\right)-\mu_{\Im\left(\mathbf{d}\right)})}{\sqrt{2\sigma^2_{\Im\left(\mathbf{d}\right)}}}\right] \hspace{-1.4mm}\sqrt{\frac{c_\sigma\mu_{\mathbf{v}^{\{L\}}}}{2R}}.
\end{equation}

\section{Results}
\label{sec:results}


For the sake of simplification, in the proposed approach the parameters $c_\sigma$ and $\mu_{\mathbf{v}^{\{l\}}}$ were set to $1$, for all layers. Then, the initializations and normalizations become
\begin{equation}
    \mathbf{b}^{\{l\}}=\mathbf{0}+\jmath\mathbf{0},
\end{equation}
\begin{equation}
    \boldsymbol{\upsigma}^{\{l\}}=\mathbf{1}+\jmath\mathbf{1},
\end{equation}
\begin{equation}
    \boldsymbol{\Gamma}^{\{l\}} \sim \mathcal{C}\mathcal{G}\left(0,\frac{1}{2O^{\{l-1\}}} \right),
\end{equation}
\begin{equation}
    \mathbf{W}^{\{l\}} \sim \mathcal{CG}\left( 0,\frac{5 O^{\{l-1\}}\exp (2 )}{6\mathrm{I^{\{l-1\}}} O^{\{l\}}}\right) ,
\end{equation}
\begin{equation}
    \mathbf{\bar{x}} = \frac{(\mathbf{x}-\mu_{\mathbf{x}})}{\sqrt{\sigma^2_{\mathbf{x}}}} \sqrt{\frac{1}{2P}},
\end{equation}
\begin{equation}
    \mathbf{\bar{d}} = \frac{(\mathbf{d}-\mu_{\mathbf{d}})}{\sqrt{\sigma^2_{\mathbf{d}}}} \sqrt{\frac{1}{2R}}.
\end{equation}

For the random initialization, we defined $\sigma^2_{\boldsymbol{\Gamma}^{\{l\}}}=1$. The $K$-means and constellation-based initializations are obtained from the input and output datasets, respectively~(see \cite{Soares2024}).

Based on \cite{Soares2023}, we consider a space-time block coding~(STBC) simulation system with the 3GPP TS 38.211 specification for 5G physical channels and modulation~\cite{3gpp.38.211}. The orthogonal frequency-division multiplexing~~(OFDM) is defined with 60-kHz subcarrier spacing, 256 active subcarriers, and a block-based pilot scheme. Symbols are modulated with~16-QAM and, for the multiple-input multiple-output~(MIMO) setup, 4~antennas are employed both at the transmitter and receiver. Based on the tapped delay line-A (TDL-A) from the 3GPP TR 38.901 5G channel models~\cite{3gpp.38.901}, the MIMO channel follows the TDLA from the 3GPP TR 38.104 5G radio base station transmission and reception~\cite{3gpp.38.104}. The TDLA is described with 12~taps, with varying delays from 0.0~ns to 290~ns and powers from -26.2~dB to 0~dB. A Rayleigh distribution is used to compute each sub-channel. To avoid influencing the learning curves, we do not take into account the Doppler effect, and we do not employ the inference learning techniques proposed by~\cite{Soares2023}. The CVNNs operate with 16 inputs and 4 outputs. The inputs are taken from the OFDM demodulator outputs, one at a time~(see \cite{Soares2023}, Fig. 1). Training and validation were performed for $3,840$ and $1,280$ instances, respectively. To assess performance, we calculated the Mean Squared Error (MSE), defined as $\text{MSE}=\frac{1}{n} \sum_{i=1}^{n} (y_i - \hat{y}_i)^2$, where $y_i$ represents the total transmitted constellation symbols over all $20$ simulations and $\hat{y}_i$ represents the respective estimated symbol from the C-RBF.

\begin{figure}[t]
    \centering    \includegraphics[width=1\linewidth]{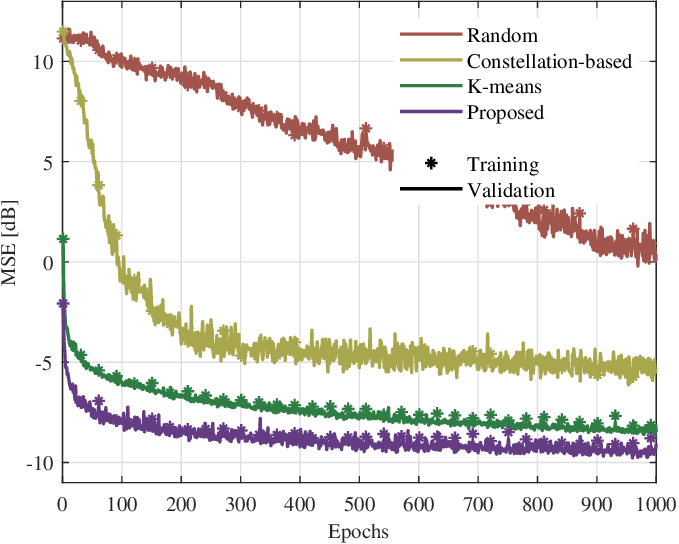}
    \caption{MSE convergence results of training~(solid lines) and validation~(asterisks) of the C-RBF initialization with a hidden layer~($I^{\{1\}}=64$~neurons) for joint channel estimation and decoding in a MIMO-OFDM $4\times4$ system, operating with 16-QAM and 256 subcarriers. Results were averaged over $20$~subsequent simulations with $\mathrm{E_b/N_0}=26$~dB. The lower the steady state MSE, the better the performance.}
    \label{fig:MIMO44_results1L}
\end{figure}

Fig.~\ref{fig:MIMO44_results1L} illustrates the mean squared error~(MSE) convergence results for $1000$~epochs of training~(solid lines) and validation~(asterisks) of the C-RBF with a hidden layer~($I^{\{1\}}=64$~neurons). Results were averaged over $20$~subsequent simulations with a bit energy to noise power spectral density ratio~$\mathrm{E_b/N_0}=26$~dB. Table~\ref{tab:parameters_single} depicts the C-RBF hyperparameters empirically optimized for each initialization scheme. None of the algorithms presented under- or over-fitting. The random initialization presented the poorest convergence results, with a steady-state error of $-6$~dB. On the other hand, the constellation-based and $K$-means initializations achieved a steady-state error of $-6$~dB and $-7.5$~dB, respectively. The best results were obtained with the proposed approach, with $-9.5$~dB of steady-state error. For comparison results regarding the convergence rate, considering $\mathrm{MSE}=-5$~dB, the proposed approach reaches this mark in five training epochs, followed by the $K$-means~(11~epochs), constellations-based~(80~epochs), and random initialization~(165~epochs).
\begin{table}[t]
\centering
\renewcommand{\arraystretch}{1.3}
\begin{threeparttable}
\caption{Single layer C-RBF optimized hyperparameters.}
\label{tab:parameters_single}
\begin{tabular}{l c c c c}
\hline
Algorithm &  $\eta_w$ & $\eta_b$ & $\eta_\gamma$ & $\eta_\sigma$\\
\hline
Random  & 0.5 & 0.5 & 0.5 & 0.5\\
Constellation-based  & 0.5 & 0.5 & 0.5 & 0.5\\
$K$-means  & 0.1 & 0.1 & 0.4 & 0.2\\
Proposed Approach & 0.1 & 0.1 & 0.4 & 0.2\\
\hline
\end{tabular}
\end{threeparttable}
\end{table}

\begin{figure}[t]
    \centering
\includegraphics[width=1\linewidth]{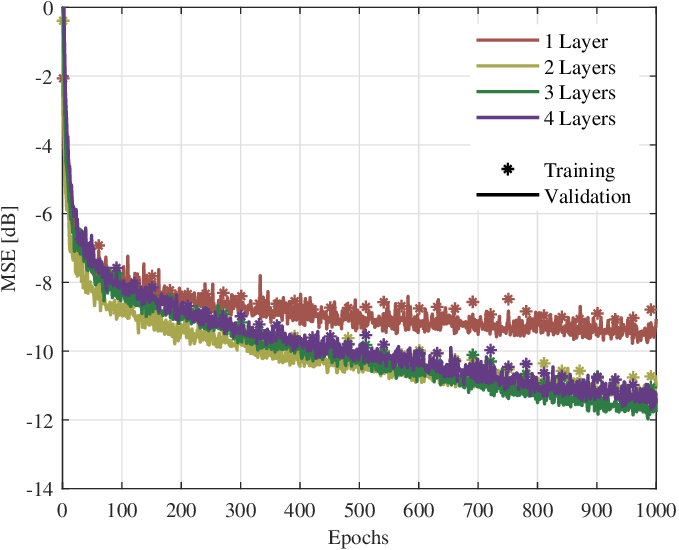}
    \caption{MSE convergence results of training~(solid lines) and validation~(stars) of the proposed initialization approach with one~($I^{\{1\}}=64$~neurons), two~($I^{\{1\}}=48$ and $I^{\{2\}}=16$~neurons), three~($I^{\{1\}}=24$, $I^{\{2\}}=24$, and $I^{\{3\}}=16$~neurons), and four~($I^{\{1\}}=16$, $I^{\{2\}}=16$, $I^{\{3\}}=16$, and $I^{\{4\}}=16$~neurons) hidden layers for joint channel estimation and decoding in a MIMO-OFDM $4\times4$ system, operating with 16-QAM and 256 subcarriers. Results were averaged over $20$~subsequent simulations with $\mathrm{E_b/N_0}=26$~dB. The lower the steady state MSE, the better the performance.}
    \label{fig:MIMO44_results2L}
\end{figure}

\begin{table}[t]
\centering
\renewcommand{\arraystretch}{1.3}
\begin{threeparttable}
\caption{Deep C-RBF optimized hyperparameters for the proposed approach.}
\label{tab:parameters_deep}
\begin{tabular}{l c c c c}
\hline
Algorithm &  $\eta_w$ & $\eta_b$ & $\eta_\gamma$ & $\eta_\sigma$\\
\hline
first layer  & 0.100 & 0.100 & 0.100 & 0.100\\
second layer & 0.050 & 0.050 & 0.050 & 0.050\\
third layer  & 0.033 & 0.033 & 0.033 & 0.033\\
fourth layer & 0.025 & 0.025 & 0.025 & 0.025\\
\hline
\end{tabular}
\begin{tablenotes}
\item These hyperparameters were optimized for the proposed initialization of the deep C-RBFs. For example, in a deep C-RBF with two hidden layers, only the first and second rows of hyperparameters are necessary. In a shallow architecture, the optimization is available in Table~\ref{tab:parameters_single}.
\end{tablenotes}
\end{threeparttable}
\end{table}

For further comparison, we have also employed the initialization schemes for C-RBFs with two, three, and four hidden layers. However, the $K$-means was not considered since it is only suitable for shallow RBFs. In addition, although several trials were attempted, no convergence was achieved for the random and constellation-based initializations. Thus, Fig.~\ref{fig:MIMO44_results2L} shows the convergence results for the proposed approach for the C-RBFs with one~($I^{\{1\}}=64$~neurons), two~($I^{\{1\}}=48$ and $I^{\{2\}}=16$~neurons), three~($I^{\{1\}}=24$, $I^{\{2\}}=24$, and $I^{\{3\}}=16$~neurons), and four~($I^{\{1\}}=16$, $I^{\{2\}}=16$, $I^{\{3\}}=16$, and $I^{\{4\}}=16$~neurons) hidden layers\footnote{For the sake of comparison, we chose a total number of neurons $N_T=64$, which was split depending on the number of layers.}. Table~\ref{tab:parameters_deep} depicts the deep C-RBF hyperparameters empirically optimized for each hidden layer. Unlike the other initialization schemes, the proposed approach achieves reasonable convergence for all architectures. One may note that the steady-state MSE results converged to the same value by the multi-layered architecture. This result is due to the number of neurons utilized to create the C-RBF layers. For the layers with the lowest number of neurons performed bottlenecks, impacting results. In order to circumvent this issue, more neurons could be adopted per layer; nonetheless, it does not affect the convergence verification.
\section{Conclusion}
\label{sec:conclusion}

This paper presents an in-depth analysis of the initialization process in complex-valued radial basis function~(C-RBF) neural networks. Our findings have elucidated the intricate dependencies involved in the initialization process. Specifically, the normalization of the input and output datasets depends on the number of inputs and outputs, respectively. Furthermore, synaptic weights are influenced by the number of neurons and outputs per layer, whereas center vectors are dependent on the number of inputs per layer. Therefore, the proposed approach is robust to changes in the neural network architecture, such as the number of inputs, outputs, hidden layers, and neurons. This innovation is particularly impactful for deploying these networks in real-world scenarios, which require robustness for a wide range of different configurations with no room for ad hoc adjustments. In a carefully designed simulation environment, conforming to 3GPP TS 38 standards, our proposed deep C-RBF parameter initialization technique exhibited superior convergence performance when compared to existing methods such as random initialization, $K$-means, and constellation-based initialization. Notably, for deep C-RBF architectures, our method was the only one that achieved successful convergence, highlighting its unique efficacy and adaptability. The implications of these results are manifold. First, they introduce a robust and effective initialization method that can significantly improve the training and performance of C-RBF neural networks, particularly in challenging 5G MIMO systems. Secondly, they lay the foundation for future research, opening avenues for the exploration of adaptive initialization techniques and offering the potential for extending our framework to other neural network architectures. In future works, we plan to validate the robustness of our proposed approach through more exhaustive experiments. We also aim to explore the applicability of our initialization framework to other neural network architectures, thereby contributing to the broader advancement of neural network-based solutions in digital communications.

%

\bibliographystyle{myIEEEtran.bst}
\bibliography{references.bib}
\balance
\end{document}